\documentclass{myaa}

\usepackage[varg]{txfonts}

\usepackage{natbib}
\usepackage{graphicx}
\usepackage{epstopdf}
\usepackage{amssymb}
\usepackage{float}

\begin{document}

\title{Model of Neptune's protoplanetary disk forming in-situ its surviving regular satellites after Triton's capture and comparison of the protoplanetary disks of the four gaseous giants}

\author{Dimitris M. Christodoulou\inst{1,2}  
\and 
Demosthenes Kazanas\inst{3}
}

\institute{
Lowell Center for Space Science and Technology, University of Massachusetts Lowell, Lowell, MA, 01854, USA.\\
\and
Dept. of Mathematical Sciences, Univ. of Massachusetts Lowell, 
Lowell, MA, 01854, USA. \\ E-mail: dimitris\_christodoulou@uml.edu\\
\and
NASA/GSFC, Laboratory for High-Energy Astrophysics, Code 663, Greenbelt, MD 20771, USA. \\ E-mail: demos.kazanas@nasa.gov \\
}


\def\gsim{\mathrel{\raise.5ex\hbox{$>$}\mkern-14mu
                \lower0.6ex\hbox{$\sim$}}}

\def\lsim{\mathrel{\raise.3ex\hbox{$<$}\mkern-14mu
               \lower0.6ex\hbox{$\sim$}}}

\abstract{
We fit an isothermal oscillatory density model of Neptune's protoplanetary disk to the surviving regular satellites and its innermost ring and we determine the radial scale length of the disk, the equation of state and the central density of the primordial gas, and the rotational state of the Neptunian nebula. Neptune's regular moons suffered from the retrograde capture of Triton that disrupted the system. Some moons may have been ejected, while others may have survived inside their potential minima. For this reason, the Neptunian nebula does not look like any of the nebulae that we modeled previously. In particular, there must be two density maxima deep inside the core of the nebula where no moons or rings are found nowadays. Even with this strong assumption, the recent discovery of the minor moon N XIV complicates further the modeling effort. With some additional assumptions, the Neptunian nebula still shares many similarities with the Uranian nebula, as was expected from the relative proximity and similar physical conditions of the two systems. For Neptune's primordial disk, we find a steep power-law index ($k=-3.0$), needed to accommodate the arrangement of the outer moons Larissa, N XIV, and Proteus. The rotation parameter that measures centrifugal support against self-gravity is quite small ($\beta_0=0.00808$), as is its radial scale length (13.6 km). The extent of the disk ($R_{\rm max}=0.12$ Gm) is a lot smaller than that of Uranus ($R_{\rm max}=0.60$ Gm) and Triton appears to be responsible for the truncation of the disk. The central density of the compact Neptunian core and its angular velocity are higher than but comparable to those of Uranus' core. In the end, we compare the models of the protoplanetary disks of the four gaseous giants.}
\keywords{planets and satellites: dynamical evolution and stability---planets and satellites: formation---protoplanetary disks}

\authorrunning{ }
\titlerunning{Formation of surviving Neptunian satellites}

\maketitle


\section{Introduction}\label{intro}

In previous work \citep{chr19a,chr19b,chr19c,chr19d}, we presented isothermal models of the solar, Jovian, Uranian, and Saturnian primordial nebulae capable of forming protoplanets and, respectively, protosatellites long before the central object is actually formed by accretion processes. This entirely new ``bottom-up'' formation scenario is currently observed in real time by the latest high-resolution ($\sim$1-5~AU) observations of many protostellar disks by the ALMA telescope \citep{alm15,and16,rua17,lee17,lee18,mac18,ave18,cla18,kep18,guz18,ise18,zha18,dul18,fav18,har18,hua18,per18,kud18,lon18,pin18,vdm19}.   In this work, we apply the same model to Neptune's primordial disk in which seven regular satellites appear to have survived after the retrograde capture of the large moon Triton and the dynamical interactions that ensued thereafter.

No model produces an acceptable fit to the present-day arrangement of the regular moons of Neptune, which immediately lends support to previous speculations that this system was severely disturbed by Triton's capture \citep{jac04,agn06,jac09}. The following problems combined are responsible for invalidating all of our standard nebular models:
\begin{itemize}
\item[(a)] The Neptunian disk appears to be extremely small; the semimajor axis of the outermost and largest regular moon Proteus is only 0.12 Gm. Triton was captured at a nearby orbit of 0.35 Gm and it is quite possible that it truncated the original disk.
\item[(b)] The innermost three moons orbit in an extremely tight configuration (radial width of only 4300 km) and no model can place them to separate density maxima.
\item[(c)] The Galle ring is the nearest structure to the planet at a mean radius of 0.0419 Gm (compare this radius to the semimajor axis of the innermost moon Naiad orbiting at 0.0482 Gm). No model can generate its first density maximum that far out in the core, so we believe that originally there were more sites of moon and/or ring formation, much deeper inside the core of the nebula, that are currently empty.
\end{itemize}
Given these peculiarities, we attempted to optimize a somewhat more sophisticated Neptunian model which however carries additional assumptions about the state of the nebula after the capture of Triton:
\begin{itemize}
\item[(a)] We fit a small primordial disk that reaches out only to the orbit of Proteus ($R_{\rm max}=0.12$ Gm), but we also check for additional density maxima just beyond $R_{\rm max}$.
\item[(b)] We assign the three innermost moons to one density peak. We retain Thalassa (with a semimajor axis of 0.05 Gm), the moon in the middle, to represent that single peak.
\item[(c)] We assign the Galle ring to the density peak interior to that of Thalassa, but this cannot be the innermost location of the arrangement. We are forced to leave the first two density peaks empty, so that the next outward peak may reach out to the location of the Galle ring.
\end{itemize}
With these additional assumptions, a good-quality optimized model can be found. We describe this model in \S~\ref{models2}. We also derive the physical properties of such a Neptunian disk and we compare them to the properties that we have obtained for the disks of the other gaseous giants, in particular of the disk of the neighboring Uranus. In \S~\ref{disc}, we summarize and discuss our results; and we compare the four protoplanetary disks of the gaseous giants that we have modeled in our current effort.

\begin{figure}
\begin{center}
    \leavevmode
      \includegraphics[trim=0.2 0.2cm 0.2 0.2cm, clip, angle=0, width=10 cm]{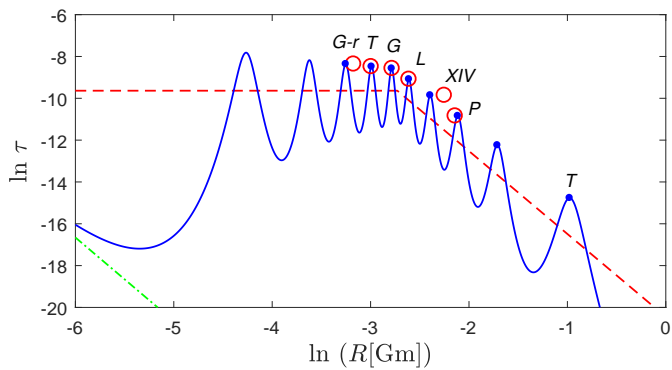}
      \caption{Equilibrium density profile for the midplane of Neptune's primordial protoplanetary disk that formed its rings and regular moons. The center of the Galle ring was also included. (Key: G-r:Galle-ring, T:Thalassa, G:Galatea, L:Larissa, XIV:N XIV, P:Proteus, T:Triton.) The best-fit parameters are $k=-3.0$, $\beta_0=0.00808$, and $R_1=0.0653$~Gm. The radial scale length of the disk is $R_0=13.6$~km. The Cauchy solution (solid line) has been fitted to the present-day regular moons of Neptune (and the center of the Galle ring) so that its density maxima (dots) correspond to the observed semimajor axes of the orbits of the moons (open circles). The two innermost density maxima (at 0.0140 Gm and 0.0267 Gm) cannot be fitted and they are left empty. The density maximum corresponding to the location of Larissa was scaled to a distance of $R_L=0.07355$~Gm. The mean relative error of the fit is 7.8\%, affirming that this model produces a good match to the observed data points under the additional assumptions stated in \S~\ref{intro}. The intrinsic analytical solution (dashed line) and the nonrotating analytical solution (dash-dotted line) are also shown for reference. 
\label{fig1}}
  \end{center}
\end{figure}

\section{Physical Model of Neptune's Protoplanetary Disk}\label{models2}

\subsection{Best-Fit Neptunian disk model}\label{model1}


In this modeling effort, we used three free parameters ($k$, $\beta_0$, and $R_1$) to fit the current orbits of the Galle ring and five of the regular moons, leaving however the first two density peaks empty (Fig.~\ref{fig1}). These two peaks are located at 0.0140 Gm and 0.0267 Gm, respectively, in the best-fit model.\footnote{The first peak is located inside the current planet and the second peak is barely outside. This is sufficient to justify the absence of moonlets, although the second peak could have hosted a ring, just like Saturn hosts its D ring very near its surface.} The third peak located at 0.0387 Gm comes close to matching the orbit of the Galle ring. 

We find the following physical parameters from the best-fit model: $k=-3.0$, $\beta_0=0.00808$, and $R_1=0.0653$ Gm (close to the orbit of Galatea). The radial scale of the model was determined by fitting the density peak that corresponds to the orbit of Larissa to its distance of 0.07355 Gm, and the scale length of the disk then turns out to be extremely small, i.e., $R_0=13.6$ km. This model is certainly stable to nonaxisymmetric self-gravitating instabilities because of the low value of $\beta_0$ \citep[the critical value for the onset of dynamical instabilities is $\beta_*\simeq 0.50$;][]{chr95}.

An outer flat-density region beyond a radius of $R_2$ did not affect the fits and it was dropped from the list of free parameters. This means that we cannot determine with certainty any properties for the outer disk that may have been truncated during Triton's capture. Nevertheless, the best-fit model shows two density peaks beyond the orbit of Proteus that are located in Fig.~\ref{fig1} at 0.1806 Gm ($\ln R\approx -1.7$) and 0.3764 Gm ($\ln R\approx -1.0$), respectively. The former peak is presently empty, whereas the latter peak is occupied by Triton and indicates that this moon was captured inside the outer protoplanetary disk, destroying all structures on and near its path (see the schematic arrangement of Neptune's moons, together with Triton, in Fig.~\ref{fig_planets} below). On the other hand, Triton does not appear to have accreted substantial local material in its run through the outer disk because its composition is similar to that of Pluto and the other objects in the Kuiper belt \citep{agn06}.

\subsection{Physical parameters from the best-fit Neptunian model}\label{rhomax1}

Using the scale length of the disk $R_0$ and the definition $R_0^2 = c_0^2/(4\pi G\rho_0)$, we write the equation of state for the Neptunian circumplanetary gas as
\begin{equation}
\frac{c_0^2}{\rho_0} \ = \ 4\pi G R_0^2 \ = \ 1.56\times 10^{6} 
{\rm ~cm}^5 {\rm ~g}^{-1} {\rm ~s}^{-2}\, ,
\label{crho1}
\end{equation}
where $c_0$ and $\rho_0$ are the local sound speed and the local density in the inner disk, respectively, and $G$ is the gravitational constant.
For an isothermal gas at temperature $T$, ~$c_0^2 = {\cal R} T/\overline{\mu}$, where $\overline{\mu}$ is the mean molecular weight and ${\cal R}$ is the
universal gas constant. Hence, eq.~(\ref{crho1}) can be rewritten as
\begin{equation}
\rho_0 \ = \ 53.4 \,\left(\frac{T}{\overline{\mu}}\right) \
{\rm ~g} {\rm ~cm}^{-3}\, ,
\label{trho1}
\end{equation}
where $T$ and $\overline{\mu}$ are measured in degrees Kelvin and 
${\rm ~g} {\rm ~mol}^{-1}$, respectively. 

For the coldest gas with $T \geq 10$~K 
and $\overline{\mu} = 2.34 {\rm ~g} {\rm ~mol}^{-1}$ (molecular hydrogen and
neutral helium with fractional abundances $X=0.70$ and $Y=0.28$ by
mass, respectively), we find that
\begin{equation}
\rho_0 \ \geq \ 228 \ {\rm ~g} {\rm ~cm}^{-3}\, .
\label{therho1}
\end{equation}

Using the above characteristic density $\rho_0$ of the inner disk
in the definition of ~$\Omega_J\equiv\sqrt{2\pi G\rho_0}$, we determine the Jeans frequency (i.e., the weight) of the self-gravitating disk:
\begin{equation}
\Omega_J \ = \ 9.8\times 10^{-3} {\rm ~rad} {\rm ~s}^{-1}\, .
\label{thej1}
\end{equation}
Then, using the model's value $\beta_0 = 0.00808$ in the definition 
of ~$\beta_0\equiv \Omega_0 /\Omega_J$, we determine the angular velocity of the uniformly-rotating core ($R_1\leq 0.0653$~Gm), viz.
\begin{equation}
\Omega_0 \ = \ 7.9\times 10^{-5} {\rm ~rad} {\rm ~s}^{-1}\, .
\label{theom1}
\end{equation}
For reference, this value of $\Omega_0$ for the core of the Neptunian nebula corresponds to an orbital period of $P_0=0.92$~d. This value is close to the present-day orbital period of N XIV (0.936 d) that is orbiting in-between the orbits of the two largest regular moons, Proteus and Larissa.
So, as with the disks of the other gaseous giants, this value of $P_0$ lands in the region where the largest regular moons were formed and survived  in the Neptunian nebula.

\begin{table*}
\caption{Comparison of the protoplanetary disks of Jupiter, Saturn, Uranus, and Neptune}
\label{table1}
\begin{tabular}{llllll}
\hline
Property & Property & Jupiter's  & Saturn's &Uranus'& Neptune's \\
Name     & Symbol (Unit) & Model 2 & Model &Model& Model \\
\hline
Density power-law index & $k$  &   $-1.4$  	       &  $-4.5$ &$-0.96$&  $-3.0$\\

Rotational parameter & $\beta_0$  &    0.0295    &  0.0431 &0.00507&  0.00808 \\

Inner core radius & $R_1$ (Gm)  &   0.220     &  0.321 &0.0967&  0.0653 \\

Outer flat-density radius & $R_2$ (Gm)  &   5.37     & 1.21 &$\cdots$&  $\cdots$ \\

Radial extent of the density power law & $\Delta R$ (Gm) & 5   & 0.9 &$\cdots$&  $\cdots$ \\

Scale length & $R_0$ (km) &   368     & 395 &27.6& 13.6  \\

Equation of state & $c_0^2/\rho_0$ (${\rm cm}^5 {\rm ~g}^{-1} {\rm ~s}^{-2}$) & $1.14\times 10^9$ &    $1.31\times 10^9$ &$6.39\times 10^6$&  $1.56\times 10^6$ \\

Minimum core density for $T=10$~K, $\overline{\mu} = 2.34$ & $\rho_0$ (g~cm$^{-3}$)         &    0.31   			    & 0.27 &55.6& 228 \\

Isothermal sound speed for $T=10$~K, $\overline{\mu} = 2.34$ & $c_0$ (m~s$^{-1}$) & 188 & 188  & 188  & 188 \\

Jeans gravitational frequency & $\Omega_J$ (rad~s$^{-1}$)    &    $3.6\times 10^{-4}$ &    $3.4\times 10^{-4}$  &$4.8\times 10^{-3}$& $9.8\times 10^{-3}$ \\

Core angular velocity & $\Omega_0$ (rad~s$^{-1}$)    &    $1.1\times 10^{-5}$ 	   &   $1.5\times 10^{-5}$  &$2.5\times 10^{-5}$&  $7.9\times 10^{-5}$ \\

Core rotation period & $P_0$ (d)  &    6.8    &  5.0 &3.0&  0.92 \\

Maximum disk size & $R_{\rm max}$ (Gm)  &    12      & 3.6 &0.60&  0.12 \\
\hline
\end{tabular}
\end{table*}

\begin{figure}
\begin{center}
    \leavevmode
      \includegraphics[trim=0.2 1.5cm 0.2 1.5cm, clip, angle=0, width=10 cm]{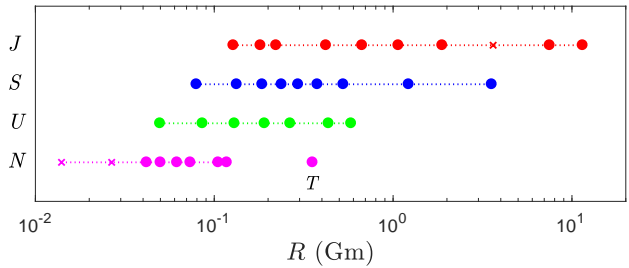}
      \caption{Schematic diagram of the protoplanetary disks of the gaseous giants that we have modeled so far. (Key: {\it J}:Jupiter, {\it S}:Saturn, {\it U}:Uranus, {\it N}:Neptune, {\it T}:Triton.) The three crosses represent the empty density peaks, predicted by the best-fit models, in which no planets or rings have been found orbiting yet. It is evident that the regular moon arrangements of Jupiter and Saturn are roughly similar. On the other hand, the arrangements of the moons of Uranus and Neptune are different, most likely because the capture of Triton ({\it T}) truncated the outer Neptunian circumplanetary disk.
\label{fig_planets}}
  \end{center}
\end{figure}

\subsection{Comparison between all best-fit models}\label{comp}

We show a comparison between the physical parameters of the best-fit models of Jupiter, Saturn, Uranus, and Neptune in Table~\ref{table1}. 
There are many striking differences between parameters of any of these models. But there are also quite a few similarities in the Jupiter-Saturn and Uranus-Neptune pairs, and pairing the models in this way makes comparisons much easier. Either one of these two pairs shows the following similarities and differences in their properties:
\begin{itemize}
\item[(a)] The paired models are similar in centrifugal support ($\beta_0$), core radius ($R_1$), disk scale length ($R_0$), equation of state ($c_0^2/\rho_0$), central density ($\rho_0$), weight ($\Omega_J$), and core angular velocity ($\Omega_0$).
\item[(b)] There are striking differences in the power-law index ($k$) and disk size ($R_{\rm max}$).
\end{itemize}
There is also a difference in $R_2$ and $\Delta R=R_2-R_1$ in the Jupiter-Saturn pair in which these parameters are meaningful. \\

The similarities can be understood as the result of the relative proximity and the sizes of the planets and their primordial disks in each pair. The differences all stem from the different sizes of the disks ($R_{\rm max}$), but these differences have radically different origins. In particular:
\begin{itemize}
\item[(a)] The disk of Jupiter was at least 3.3 times larger than Saturn's disk. Saturn's smaller disk had to form an equal number of unequally spaced regular moons within a much smaller radial extent. Thus, Saturn's disk must have had a much steeper density profile ($k=-4.5$ versus $k=-1.4$), precisely as is found in these models.
\item[(b)] The disks of Uranus and Neptune should have been comparable in size (because of their proximity within the solar nebula), if it were not for Triton whose retrograde capture changed the original Neptunian disk considerably by truncating this disk to a much smaller size (merely a size of $R_{\rm max} = 0.12$ Gm). Triton's capture in the outer disk of the primordial Neptunian nebula wreaked havoc in the arrangement of Neptune's original protosatellites (Fig.~\ref{fig_planets}), possibly ejecting some moonlets and modifying the orbits of the remaining moons that were lying inside potential minima. Most likely, the truncation of the circumplanetary disk of Neptune resulted in a much smaller disk, and this smaller disk region is where the surviving regular satellites are observed nowadays (within a tiny annular width of only 0.07 Gm). 
\end{itemize}

All models in Table~\ref{table1} exhibit high enough central densities to ensure that a ``bottom-up'' hierachical formation occurred around all of the protoplanets. As we have found for the other gaseous giants, protosatellites are seeded very early inside their nebular disks and long before the central protoplanets are fully formed; these compact moon/ring systems likely come to existence in $< 0.1$ Myr \citep{har18} and long before the central star has become fully formed \citep[see also][]{gre10}.

\section{Summary and Discussion}\label{disc}

We have constructed isothermal differentially-rotating protoplanetary models of the Neptunian nebula, the primordial disk in which the surviving regular moons and the Galle inner ring were formed (\S~\ref{models2}). The best-fit model is shown in Fig.~\ref{fig1} and its physical parameters are listed in Table~\ref{table1}, along with the physical parameters of all the other nebular disks of the gaseous giants. Neptune's best-fit model is unique in that we had to concede that the two inner density peaks are currently empty and the three innermost moons were formed within one potential minimum. Only under these conditions, can we reduce the mean relative error of the best-fit model to an acceptable value of 7.8\%. 

We have compared this special model of Neptune's disk (after Triton was captured and caused significant changes) to the best-fit model of Uranus, the nearest system to Neptune. There are many similarities between the two models and some pronounced differences as well, mainly due to the sizes of the primordial disks (Fig.~\ref{fig_planets}). The same outcome occurs in the comparison of the primordial disks of the larger two planets in the solar system, Jupiter and Saturn. The disks of Jupiter and Saturn share many similarities and the few differences can be understood as resulting from the difference in the sizes of these primordial disks.

This is the final installment of protosatellite modeling around the outer gaseous giants. What have we learned from this effort? We have settled in four fundamental facts (see Fig.~\ref{fig_planets} and Table~\ref{table1}):
\begin{enumerate}
\item[1.] The physical conditions in Jupiter and Saturn's primordial nebulae were similar, despite the fact that Jupiter's disk was at least 3.3 times larger.
\item[2.] The physical conditions around the icy giants, Uranus and Neptune, were also similar, although Neptune's outer disk suffered a severe truncation from the capture of Triton that orbited inside the outer primordial disk and wreaked havoc in the original regular protosatellite arrangement.
\item[3.] There were no significant migrations of the gaseous protoplanets or their (surviving or otherwise safely-forming) protosatellites because these objects grew safely inside local gravitational potential minima that were provided by the gaseous self-gravitating disks around the central accreting objects. 
\item[4.] The results support strongly a ``bottom-up'' hierachical formation scenario according to which moons and rings form first around their protoplanetary cores and then these cores complete their formation before the Sun is actually formed at the center of the solar nebula.
\end{enumerate}
Taken altogether, our results argue against violent formation and evolution of protoplanets and protosatellites  and, in particular, extensive migrations in the solar system such as that hypothesized in the Nice model \citep{tsi05,mor05,gom05}. Neptune's present-day moon arrangement makes yet the best case against migration/destruction/violence: even though Triton came in from the Kuiper belt, was captured by Neptune, and destroyed the outer proto-Neptunian disk in the process (probably ejecting outer moons in the process), nevertheless seven small regular satellites survived the ordeal, presumably sitting safely inside gravitational potential troughs provided by the gas and oblivious to the armageddon brought forth by that incoming large fragment. 

\end{document}